\documentclass[twocolumn,showpacs,prl,aps]{revtex4-1}
\usepackage{bm,color,amsmath,bbold,bbm,amssymb,mathrsfs,latexsym,graphicx,psfrag}

\newcommand{\bs}[1]{\boldsymbol{#1}}
\newcommand{\be}{\begin{equation}}
\newcommand{\ee}{\end{equation}}
\newcommand{\bea}{\begin{eqnarray}}
\newcommand{\eea}{\end{eqnarray}}

\renewcommand{\phi}{\varphi}
\renewcommand{\epsilon}{\varepsilon}
\renewcommand{\vec}[1]{{\bf #1}}

\usepackage{soul}
\sethlcolor{green}
\setstcolor{red}

\begin{document}

\title{Unconventional Fermi surface instabilities in the Kagome Hubbard Model}

\author{Maximilian L. Kiesel${}^1$} 
\author{Christian Platt${}^1$} 
\author{Ronny Thomale${}^{2}$}

\affiliation{${}^1$Institute for Theoretical
  Physics, University of W\"urzburg, Am Hubland, D
  97074 W\"urzburg} 
\affiliation{${}^2$ Institut de th\'eorie des ph\'enom\`enes physiques, \'Ecole Polytechnique F\'ed\'erale de Lausanne (EPFL), CH-1015 Lausanne}
\date{\today}

\begin{abstract}
We investigate the competing Fermi surface instabilities in the Kagome tight-binding model. Specifically, we consider onsite and short-range Hubbard interactions in the vicinity of van Hove filling of the dispersive Kagome bands where the Fermiology promotes the joint effect of enlarged density of states and nesting. The sublattice interference mechanism [Kiesel and Thomale, Phys. Rev. B Rapid Comm., in press.] allows us to explain the intricate interplay between ferromagnetic fluctuations and other ordering tendencies. On the basis of functional renormalization group used to obtain an adequate low-energy theory description, we discover finite angular momentum spin and charge density wave order, a two-fold degenerate $d$-wave Pomeranchuk instability, and $f$-wave superconductivity away from van Hove filling. Together, this makes the Kagome Hubbard model the prototypical scenario for several unconventional Fermi surface instabilities. 
\end{abstract}

\maketitle

{\it Introduction.} The interplay of Fermiology and interactions gives rise to a plethora of ordering phenomena in two-dimensional electron systems. Starting from an itinerant electron picture, the density of states (DOS) at low energies around the Fermi level as well as nesting features of the Fermi surface are the relevant parameters of the kinetic theory. In the limit of weak interactions imposed on the non-interacting electrons where a perturbative treatment is asymptotically exact, superconducting order as a phase-coherent superposition of Cooper pairs is the generically encountered Fermi surface instability~\cite{kohn-65prl524}. This situation can be changed in various different ways\textcolor{red}{:} By enhancing the interaction scale or via nesting and enlarged DOS at the Fermi level, order due to condensation of particle-hole pairs can become competitive and even favorable to superconductivity. Prominent examples include magnetic (charge) order via a spin (charge) density wave which can induce superconductivity as a function of doping or pressure. 

Since the discovery of the cuprates, it has widely been appreciated that electronically mediated interactions in particular, can give preference to electron condensates with finite angular momentum of the condensing pairs. It implies that the pairs which condense form at finite distance, so as to minimize Coulomb repulsion in the case of a Cooper pair. This yields a momentum dependence of the associated mean field order parameter which can impose nodes on the Fermi surface in the ordered phase. In the case of a particle-hole condensate with opposite charge of the pair constituents, it is less generic that a pair of finite angular momentum should be energetically preferable. As another difference to particle-particle pairs, the orbital angular momentum of a particle-hole pair does not unambiguously determine the spin of the pair to be singlet or triplet, which implies an even richer variety of possible orderings~\cite{schulz89prb2940,nersesyan-91jpcm3353,nayak-00prb4880}. While this is short of a complete characterization, one important factor in favoring such orders is given by a tuned arrangement of longer range interactions~\cite{schulz89prb2940}, a direction which has become recently accessible experimentally with sufficient tunability in dipolar fermion models~\cite{bhongale-12prl145301}. Still, until today the main challenge in theory has been to find bare models of interacting electrons where these phases can be found as the natural ordered state at low energies.

\begin{figure}[t]
\begin{minipage}[l]{0.99\linewidth}
 \includegraphics[width=\linewidth]{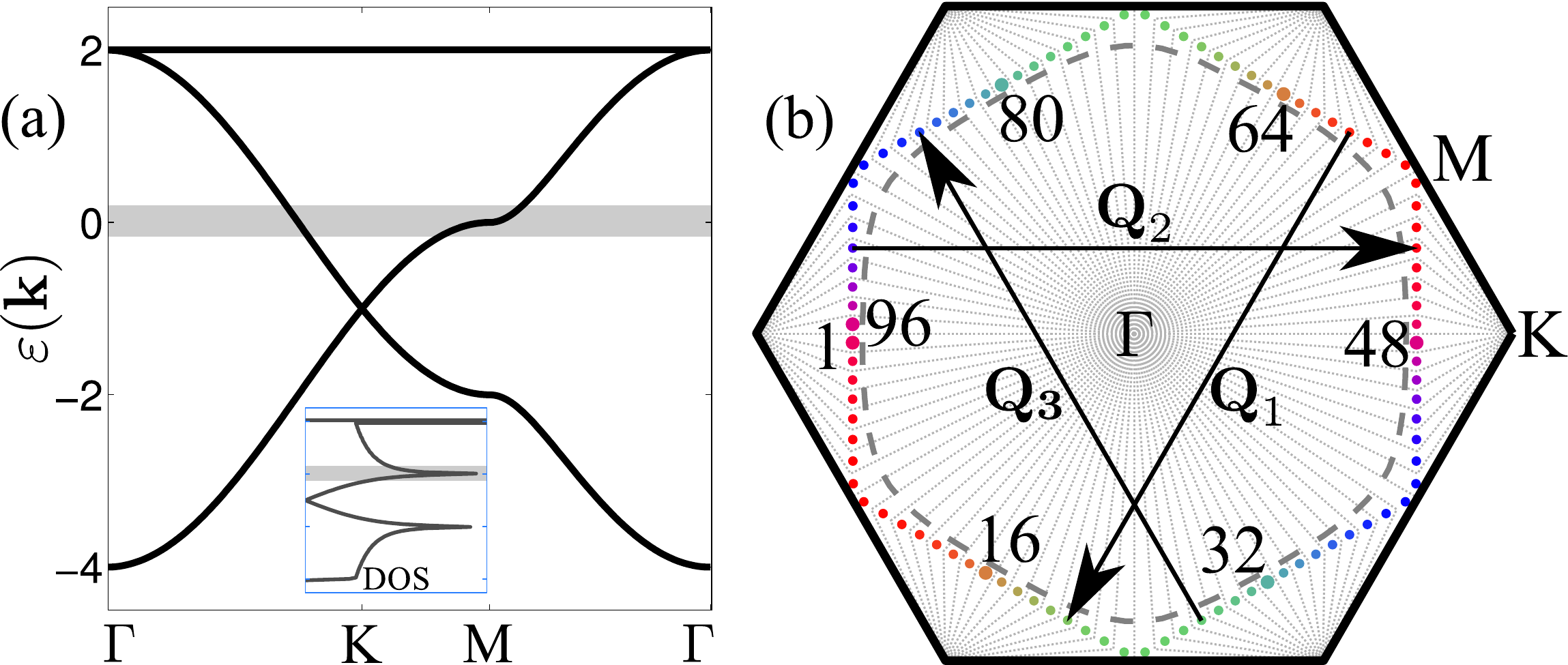}
\end{minipage}
\caption{(Color online). Fermiology of the Kagome tight-binding model. (a) Band structure where the shaded region is the vicinity around van Hove filling (VHF) $n=5/12$ with an enlarged density of states (inset). (b) Fermi surface at VHF. An $N=96$ patch discretization of the Brillouin zone is given along with the change of dominant sublattice occupation (red, blue, and green dots) on the Fermi surface. The nesting vectors $\vec{Q}_{1,2,3}$ connect different sublattice states at the Fermi level. The grey dashed line hints the Fermi surface at $n=5/12+0.02$.}
\label{fig1}
\vspace{-0pt}
\end{figure}

In this Letter, we propose and analyze the Kagome Hubbard model (KHM) as a prototypical microscopic model to realize certain kinds of such unconventional Fermi surface instabilities. The Kagome lattice~\cite{elser89prl2405} possesses a minimal three-band model due to three sites per unit cell (Fig.~\ref{fig1}a). For the KHM, the three sublattices have a fundamental impact on the preferred electronic many-body state at all regimes of coupling strength. In the strong coupling limit at half filling, the Kagome spin model exhibits strong quantum spin fluctuations and has become a primary candidate for quantum spin disordered phases~\cite{ramirez94arms,misguichlhuillier,mendels-10jpsj011001}. At intermediate coupling, comparatively little is known about the electronic quantum phases. At fillings up to the flat band (Fig.~\ref{fig1}a), ferromagnetism has been proposed on the basis of Stoner's criterion~\cite{tanaka-03prl067204}. 

For the dispersive band fillings which we address in this work, the scenario is more complicated and involves the interplay of nesting, Fermi level DOS, and interactions. This is likewise indicated from recent studies at infinitesimal coupling where the sublattice interference mechanism has been developed as a key property to understand the KHM~\cite{kiesel-arxiv1206}.   
Out of the search for microscopic material scenarios of a Kagome lattice model at intermediate coupling, the Herbertsmithites such as ${\mathrm{ZnCu}}_{3}(\mathrm{OH}{)}_{6}{\mathrm{Cl}}_{2}$ appear as an important class of candidates, while it is still hard to 
judge from neutron experiments how relevant charge fluctuations are and how the system behaves upon doping~\cite{han-12prl157202}.
Promising alternative routes start to emerge in optical Kagome lattices of ultra-cold fermionic atomic gases such as for the isotopes ${}^6\text{Li}$ and ${}^{40}\text{K}$~\cite{jo-12prl045305}. 

\begin{figure}[t]
\begin{minipage}[l]{0.99\linewidth}
 \includegraphics[width=\linewidth]{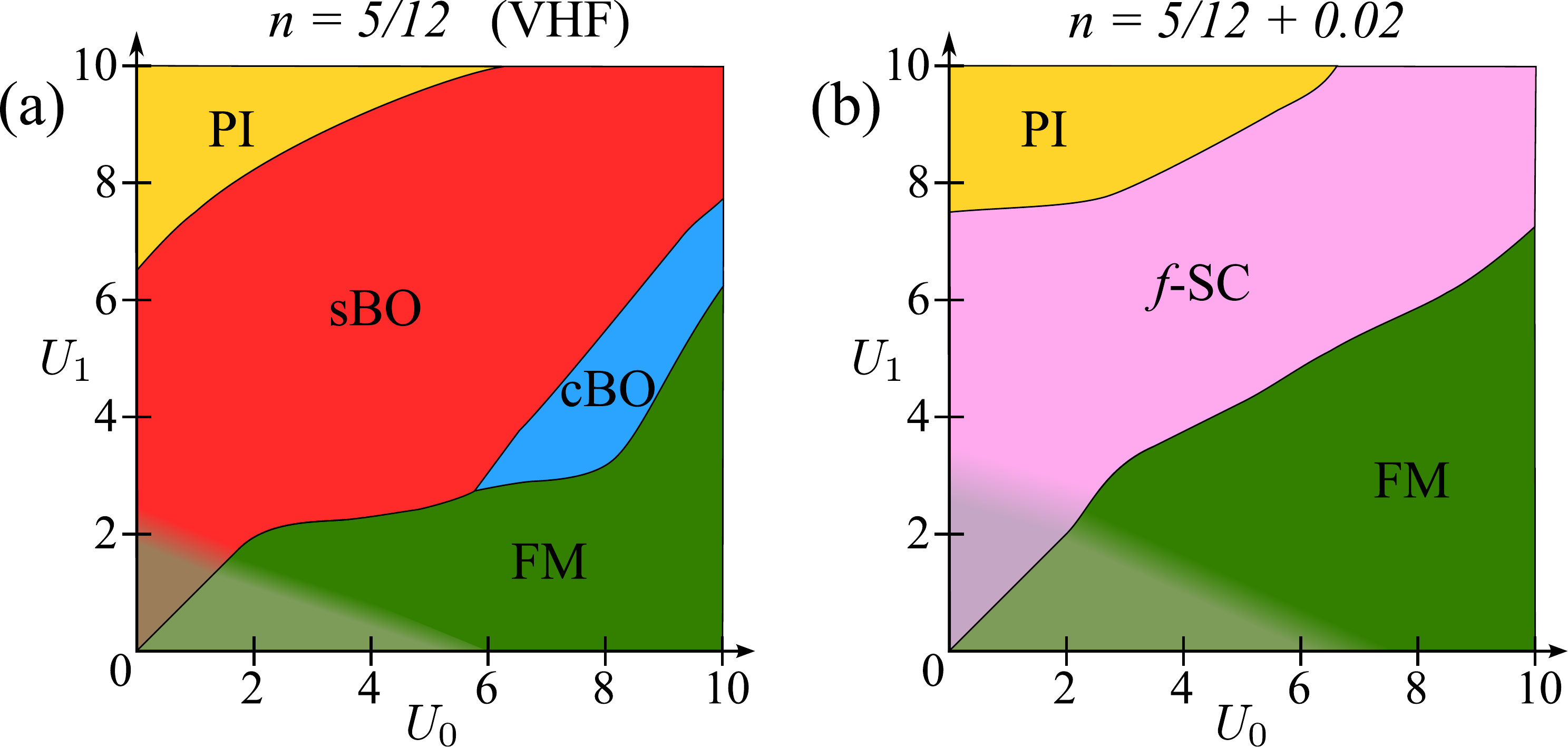}
\end{minipage}
\caption{(Color online). Phase diagram of the $U_0$-$U_1$ Kagome Hubbard model at (a) VHF $n=5/12$ and (b) away from VHF at $n=5/12+0.02$. The shaded areas for small $U_0$ and $U_1$ indicate regimes where critical ordering scales were too small to be determined. In (a) ferromagnetism (FM) is found for large $U_0$, along with a $d$-wave Pomeranchuk instability (PI) for large $U_1$ as well as intermediate spin bond order (sBO) and charge bond order (cBO) phases. In (b) the PI and FM phase persist away from VHF, while the density wave orders disappear in favor of an $f$-wave superconducting ($f$-SC) phase.}
\label{fig2}
\vspace{-0pt}
\end{figure}

\begin{figure}[t]
\begin{minipage}[l]{0.99\linewidth}
 \includegraphics[width=\linewidth]{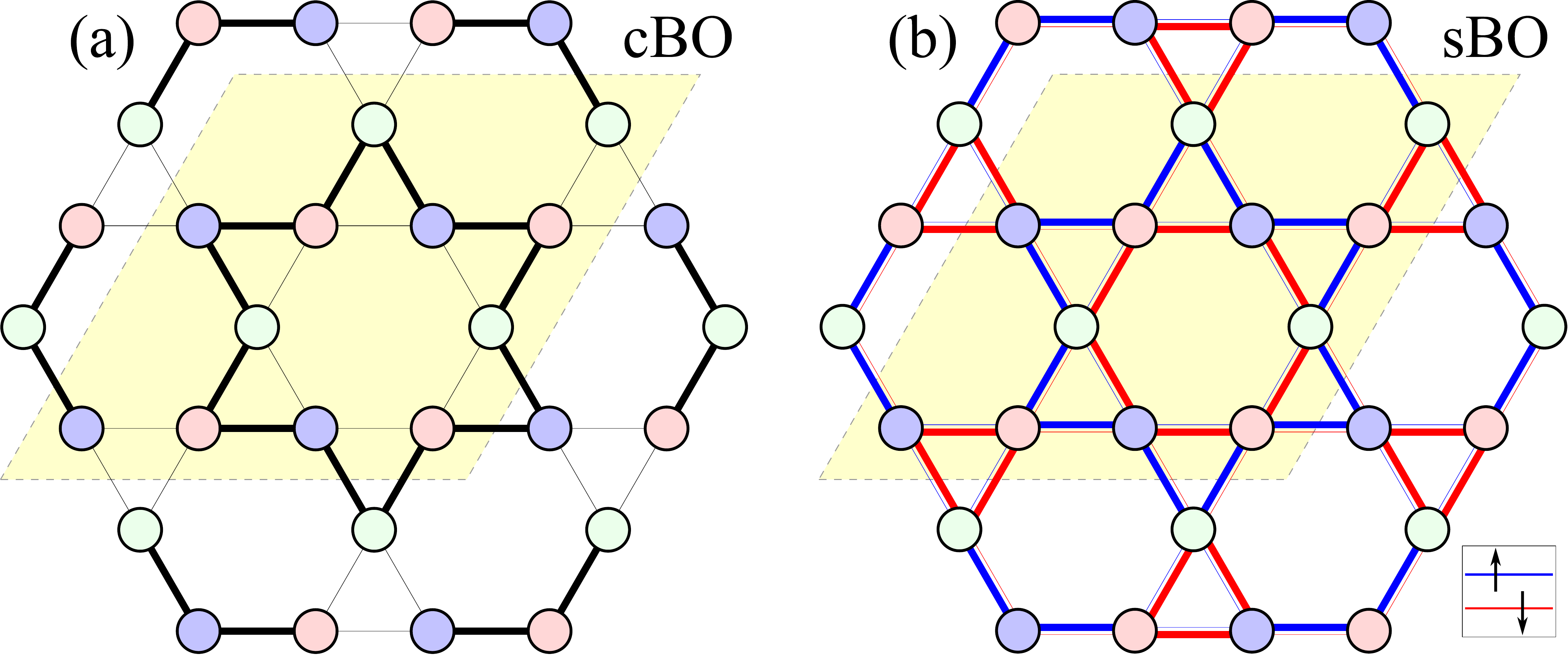}
\end{minipage}
\caption{(Color online). Real space patterns of $L=1$ density wave orders with (a) charge bond order (cBO, $S=0$) and (b) spin bond order (sBO, $S=1$). Both orders are superpositions of the the one-dimensional alternating bond orders induced along the momentum transfer $\vec{Q}_{1,2,3}$, yielding a $12$-site unit cell. Enhanced charge density bonds in (a) are indicated by thick black lines, $\uparrow$ ($\downarrow$) bonds in (b) by thick blue (red) lines.}
\label{fig3}
\vspace{-0pt}
\end{figure}

\begin{figure*}[t]
\begin{minipage}[l]{0.99\linewidth}
 \includegraphics[width=\linewidth]{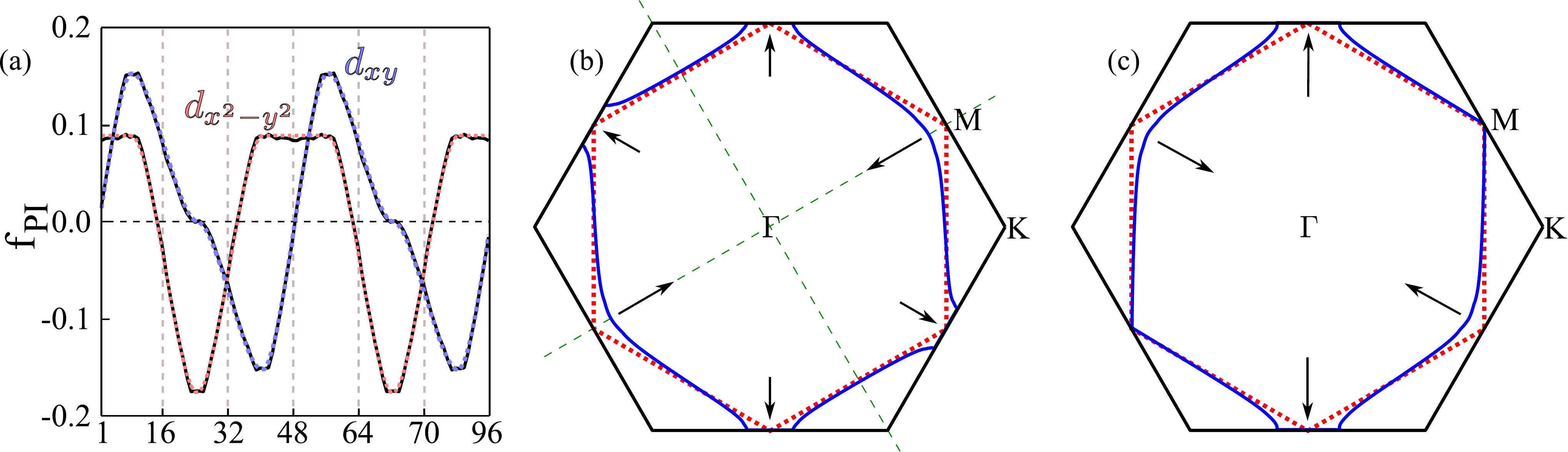}
\end{minipage}
\caption{(Color online). Pomeranchuk instability (PI) in the regime $U_1>U_0$. The $Q=0$ particle-hole pairing does not break translational symmetries and is dominated by third nearest neighbor pairing of the same Kagome sublattice which follows from the form factors of Eq.~\ref{dwaveformfacors} shown in (a). The PI lies in the $L=2$ sector ($d$-wave type) and is two-fold degenerate. It yields a distortion of the Fermi surface which can break the rotational symmetry in different ways such as to $C_{2v}$ in (b) or $C_2$ in (c).}
\label{fig4}
\vspace{-0pt}
\end{figure*}

\begin{figure}[t]
\begin{minipage}[l]{0.99\linewidth}
 \includegraphics[width=\linewidth]{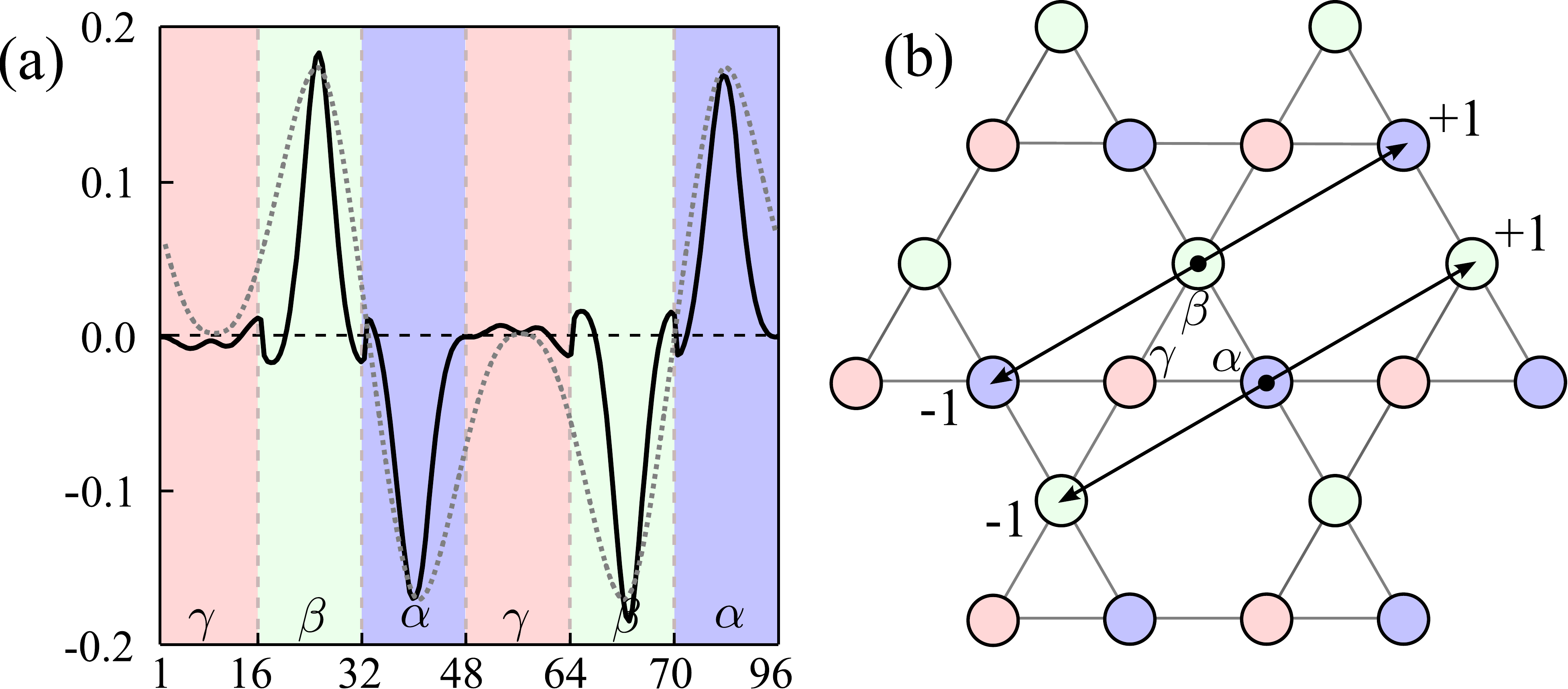}
\end{minipage}
\caption{(Color online). $f$-wave superconductivity ($f$-SC) away from VHF for $U_0 \sim U_1$. (a) $f$-wave pairing form factor (black solid line) between the $\alpha$ and $\beta$ sublattice. The signal should be small for the $\gamma$-dominated domains on the Fermi surface (red patches $1-16$ and $48-64$). The $\alpha$ and $\beta$  regimes of the form factor are best fit via second nearest neighbor pairing harmonics (dotted lines). The real space amplitudes of the $\alpha$-$\beta$ pairing is shown in (b). The total $f$-wave signal is built by adding the analogous $\alpha$-$\gamma$ and $\beta$-$\gamma$ components.}
\label{fig5}
\vspace{-0pt}
\end{figure}

{\it Main results.} As a function of local and nearest neighbor Hubbard coupling $U_0$ and $U_1$, we find a rich phase diagram of the Kagome Hubbard model (KHM) at and around van Hove filling (VHF) which is summarized in Fig.~\ref{fig2}. Right at VHF where the Fermi level DOS is maximally enhanced and nesting features of the Fermi surface are strongest (Fig.~\ref{fig1}b), the system promotes ferromagnetism (FM) for dominant $U_0$. This is a consequence of the suppression of local Hubbard matrix elements due to sublattice interference~\cite{kiesel-arxiv1206} which otherwise would give rise to spin density wave or superconducting order. As $U_1$ is enhanced, we discover a $p$-wave ($L=1$) charge bond order (cBO) and spin bond order (sBO) phase (Fig.~\ref{fig3}). This is again motivated by the sublattice interference which suppresses the energy gain of a zero angular momentum particle-hole condensate. For dominant $U_1$, we find a $d$-wave ($L=2$) Pomeranchuk instability (PI) which is two-fold degenerate due to the associated two-dimensional irreducible point group representation of the Kagome lattice, i.e. the $E_2$ element of the $C_{6v}$ symmetry group. This explains why the nematic phase resulting from there can break the rotation symmetry of the Kagome lattice in different ways, which leads to different distortions of the Fermi surface (Fig.~\ref{fig4}). As we deviate from VHF, nesting effects get reduced. While the PI (FM) phase still persist for dominant $U_1$ ($U_0$), $f$-wave superconductivity ($f$-SC) emerges in the intermediate regime as a consequence of longer range interactions and ferromagnetic fluctuations which promote spin alignment of the Cooper pairs (Fig.~\ref{fig5}).

{\it Kagome Hubbard model.} We consider the Hamiltonian
\begin{eqnarray}
H&=&\sum_{\langle i, j \rangle} \sum_{\sigma} \left( c_{i\sigma}^\dagger c_{j\sigma}^{\phantom{\dagger}} + \text{h.c.} \right) + \mu\sum_{i,\sigma} n_{i,\sigma} \nonumber \\
&&+ U_0 \sum_{i}n_{i,\uparrow} n_{i,\downarrow} +\frac{U_1}{2}\sum_{\langle  i, j\rangle, \sigma, \sigma'} n_{i,\sigma} n_{j,\sigma'},
\label{ham}
   \end{eqnarray}
where $U_0$ denotes the local and $U_1$ the nearest neighbor Hubbard term, $\mu$ is the chemical potential, and the tight binding hopping term has been set to unity. In what follows, we adjust the chemical potential such that we are around VHF $n=5/12$, and set the Fermi level at VHF to zero energy (Fig.~\ref{fig1}a). There, the Fermi surface possesses a hexagonal form where the van Hove points are located at the $M$ point in the Brillouin zone (Fig.~\ref{fig1}b). It follows that there are three important nesting vectors $\vec{Q}_1=\pi(-\frac{1}{2},-\frac{\sqrt{3}}{2})$, $\vec{Q}_2=\pi(1,0)$, and $\vec{Q}_3=\pi (-\frac{1}{2},\frac{\sqrt{3}}{2})$, with the lattice constant between two adjacent sites set to unity. Away from VHF, the Fermi surface is rounded (gray dashed line in Fig.~\ref{fig1}b) and nesting is reduced. The nesting vectors $\vec{Q}_{1,2,3}$ connect parts of the Fermi surface whose states are dominated by different sublattices (Fig.~\ref{fig1}b). As a consequence, by transforming the local Hubbard term $U_0$ into band space, the matrix elements along the nesting vectors are suppressed which we call sublattice interference~\cite{kiesel-arxiv1206}. If this interference mechanism were absent or neglected, the short-range KHM would take on phases dictated only by the nesting vectors $\vec{Q}_{1,2,3}$, such as a spin density wave state, and $d$-wave superconductivity below some finite coupling strength~\cite{yu-12prb144402}.  Instead, the sublattice interference enhances the relevance of ferromagnetic fluctuations stemming from the local Hubbard term, and also promotes the relevance of the nearest neighbor Hubbard coupling.

{\it $N$-patch functional renormalization group.} We employ the functional renormalization group (FRG) to obtain an effective low energy description of the bare model in~\eqref{ham}, which has proved suitable in a variety of interacting two-dimensional electron systems~\cite{metzner-12rmp299}. We study how the 2-particle vertex evolves under integrating out high-energy fermionic modes along a temperature-flow cutoff scheme: $V_\Lambda(\bs{k}_1,\bs{k}_2;\bs{k}_3 ,\bs{k}_4)c_{\bs{k}_4,s}^\dagger c_{\bs{k}_3,\bar{s}}^\dagger c_{\bs{k}_2,\bar{s}}^{\phantom{\dagger}} c_{\bs{k}_1,s}^{\phantom{\dagger}} $, where $\Lambda$ is the IR cutoff approaching the Fermi surface within the flow, $\bs{k}_1$, $\bs{k}_2$ ($\bs{k}_3$, $\bs{k}_4$) denote the ingoing (outgoing) fermionic momenta, and $s,\bar{s}$ take on opposite spin orientations. This is sufficient because for a spin-rotation invariant model, the $S^z=0$ sector of the scattering vertex allows us to extract the (triplet) singlet channel by (anti-)sym\-metrization of the vertex. We neglect the self energy corrections along the flow and discretize the $\bs{k}$'s to represent specific patches in the Brillouin zone. In Fig.~\ref{fig1}b, we have depicted such a patching scheme for $N=96$ patches. The phase diagrams in Fig.~\ref{fig2} have been obtained with this level of discretization. In order to assure that the discretization yields converged results, we have employed supercomputer facilities to compute selected points of the phase diagram for up to $N=384$ patches, which corresponds to solving a $5.7\times10^7$-dimensional system of integro-differential equations. The leading diverging channel of the 2-particle vertex signals the occurrence of a Fermi surface instability. It is then decomposed into eigenmodes to obtain the associated form factor~\cite{zhai-09prb064517}.

{\it Ferromagnetism (FM).} As alluded to above, the sublattice mechanism suppresses strong finite momentum scattering channels originating from $U_0$. When the local Hubbard interaction is dominant, the large Fermi level DOS  drives ferromagnetic fluctuations without any other competing channel, which thus explains the formation of FM order. This holds at and around VHF (Fig.~\ref{fig2}). The order parameter reads $\bs{\mathcal{O}}_\text{FM}= \sum_{\bs{k},l,s,s'} \langle c_{\bs{k}l s}^\dagger \bs{\sigma}_{ss'} c_{\bs{k}l s'}^{\phantom{\dagger}} \rangle,$
where $\bs{\sigma}$  denotes the vector of Pauli matrices. In addition, the propensity towards spin alignment from the ferromagnetic fluctuation background at high energies provides further bias for the sBO phase at VHF and the $f$-wave SC phase away from VHF.

{\it Spin and charge bond order (sBO and cBO).} As $U_1$ is enhanced at VHF, the nesting vectors $\vec{Q}_{1,2,3}$ become important again. This is because the sublattice interference does not apply to nearest neighbor Hubbard interactions which connect different Kagome sublattices. As a consequence, the KHM exhibits cBO also known as the Peierls phase which is characterized as a $L=1$, $S=0$ particle-hole condensate as well as its spinful condensate counterpart $L=1$, $S=1$, i.e., the sBO phase (Fig.~\ref{fig3}). The cBO shows an alternating sequence of bonds with enhanced and reduced charge. Accordingly, the sBO exhibits alternating bonds of enhanced $\uparrow$ spin occupancy and enhanced $\downarrow$ spin occupancy. Both cBO and sBO can be deconstructed into separate bond orders along the three individual directions in the Kagome lattice, which, recast in momentum space, are exactly $\vec{Q}_{1,2,3}$. For sBO, the order parameter is given by
\begin{equation}
\bs{\mathcal{O}}_\text{sBO}= \sum \limits_{\genfrac{}{}{0pt}{}{\bs{k},s,s'}{l,m,n}} \left\langle c_{\bs{k}l s}^\dagger \bs{\sigma}_{ss'} c_{\bs{k}+\bs{Q}_m n s'}^{\phantom{\dagger}} \right\rangle \times \sin\left( \frac{\bs{Q}_m\bs{k}}{\pi} \right) \left| \epsilon_{lmn} \right|,\label{bo}
\end{equation}
where $\epsilon_{lmn}$ is the Levi-Civita tensor. Eq.~\ref{bo} takes the similar form for cBO, where $\bs{\sigma}_{ss'}$ is replaced by the identity matrix $\mathbbm{1}_{ss'}$. There are in principle three independent bond order mean fields associated with these directions. As they are independent and degenerate, however, the system gains energy by forming all of them at the same time, which results in a $12$-site unit depicted in Fig\textcolor{red}{.}~\ref{fig3}a for cBO and in Fig.~\ref{fig3}b for sBO. We checked in a $12$-band mean field free energy analysis that the system linearly gains energy from forming the individual mean fields, i.e. the ordering along the individual bond directions is independent.   

{\it Pomeranchuk instability (PI).} The PI forms for dominant $U_1$ at and around VHF. The singularity in the particle-hole channel is located at momentum transfer $Q=0$, implying that it does not break translational symmetries, but instead drives the system into a nematic phase as it breaks the lattice rotational symmetries of the KHM. The particle-hole condensate is characterized by $L=2$, $S=0$. According to the irreducible lattice representations of the Kagome lattice, a $d$-wave instability necessitates a two-fold degenerate $d_{x^2-y^2}$, $d_{xy}$ subspace of solutions~\cite{kiesel-12prb020507}. These form factors are depicted in Fig.~\ref{fig4}a. It turns out that the most dominant harmonics in this symmetry sector relate to the third nearest neighbor, i.e. equal sublattice particle-hole pairing. It is revealing to investigate what kind of Fermi surface distortion can result from such an instability. As stated before, the PI effectively generates a third nearest neighbor hybridization. Its lattice rotational character, however, is not uniquely specified because we could in principle form any real superposition of the $d_{x^2-y^2}$, $d_{xy}$ solutions~\cite{maharaj}, i.e.
\begin{eqnarray}
f_{d_{x^2-y^2}}\left(\bs{k}\right) &=& \cos\left(2k_x\right) - \cos\left(k_x\right) \cos\left(\sqrt{3} k_y\right),\nonumber\\
f_{d_{xy}}\left(\bs{k}\right) &=& \sqrt{3} \sin\left(k_x\right) \sin\left(\sqrt{3} k_y\right) \ . \label{dwaveformfacors}
\end{eqnarray}
Within our approach, we are short of an answer which linear combination is energetically preferable, since the lack of self energy damping results in an unbound energy gain from Fermi surface distortion. This ambiguity is also visible in the order parameter
\begin{equation}
\bs{\mathcal{O}}_\text{PI}= \sum_{\bs{k}, l, s} \left\langle c_{\bs{k}l s}^\dagger c_{\bs{k}l s}^{\phantom{\dagger}} \right\rangle \left(A f_{d_{x^2-y^2}}\left(\bs{k}\right) + B f_{d_{xy}}\left(\bs{k}\right) \right),
\end{equation}
where $A$ and $B$ specify the superposition.
We note, however, that the remainder rotational group down to which the PI can establish a nematic phase depends on the chosen linear combination. In Fig.~\ref{fig4}b, the remainder group is $C_{2v}$ and the Fermi surface is shifted away from all van Hove points, while for example in Fig.~\ref{fig4}c one van Hove point remains unaltered and the remainder group is $C_2$.

{\it $f$-wave superconductivity ($f$-SC).} Away from VHF for intermediate $U_0$ and $U_1$ we find an $f$-wave, i.e. $L=3$, $S=1$ Cooper pair condensate in the KHM. Fig.~\ref{fig5}a shows the form factor associated with the order and Fig.~\ref{fig5}b depicts the real space amplitude pattern which is essential to identify the dominant harmonics in the $f$-wave symmetry sector, i.e. the typical pairing distance~\cite{kiesel-12prb020507}. We find that SC pairing forms between second nearest neighbors to avoid the short-range Hubbard interaction. Denoting the three sublattices by $\alpha$, $\beta$, and $\gamma$, we can decompose the pairing into three different form factor contributions given by 
$f_{\alpha, \beta}(\bs{k})=\sin (\frac{3}{2}k_x+\frac{\sqrt{3}}{2}k_y)$, $f_{\beta, \gamma}(\bs{k})=\sin (\frac{3}{2}k_x-\frac{\sqrt{3}}{2} k_y)$, and $f_{\alpha, \gamma}(\bs{k})=\sin (\sqrt{3}k_y)$, where $f_{m,n} = f_{n,m}$ and $f_{m,m}=0$. Together, they form the order parameter
\begin{equation}
\bs{\mathcal{O}}_\text{$f$-SC}= \sum \limits_{\bs{k},m,n} \left\langle c_{\bs{k}m\uparrow}^\dagger c_{\bs{k}n\downarrow}^\dagger + c_{\bs{k}m\downarrow}^\dagger c_{\bs{k}n\uparrow}^\dagger \right\rangle f_{m,n}\left(\bs{k}\right),
\end{equation}
which can be rotated within the triplet state to the $S^z=\pm1$ sectors. Such a multi-sublattice scenario is similar to a multi-orbital superconductor where the form factor can be similarly decomposed into different orbital pairs~\cite{thomale-11prl187003}.

{\it Note added.} Upon completing the manuscript, we became aware of an independent work that investigates the Kagome Hubbard model strictly at van Hove filling through singular mode FRG, which is a complementary approach to $N$-patch FRG with enhanced radial resolution of the RG flow~\cite{wang-arxiv1208}. The discrepancies in terms of qualitative results could trace back to assumptions made in the singular mode FRG approach where the vertices are reduced to their bosonic transverse momentum dependence in the respective channel~\cite{husemann-09prb195125}. This approximation might be less valid when ordering of finite distance pairs emerges yielding important dependencies of all vertex momenta. This is the generic case for most Kagome lattice orders we find such as PI (third nearest neighbor) or $f$-wave SC (second nearest neighbor).

\begin{acknowledgments}
RT thanks A.~Maharaj and S.~Raghu for collaborations on related topics as well as A.~Eberlein and C.~Honerkamp for discussions on singular mode FRG. MK and CP are supported by DFG-FOR 1162. MK, CP, and RT acknowledge support by SPP-1458/1.
\end{acknowledgments}


\end{document}